# Lower Critical Field Measurement System based on Third-Harmonic Method for Superconducting RF Materials

Hayato Ito[a,*], Hitoshi Hayano[b], Takayuki Kubo[a,b], and Takayuki Saeki[a,b],

[a]SOKENDAI, 1-1 Oho, Tsukuba, Ibaraki 305-0801, Japan

[b]KEK: High Energy Accelerator Research Organization, 1-1 Oho, Tsukuba, Ibaraki 305-0801, Japan

[*]Corresponding author.

Tel: +81-29-864-5200

E-mail address: hayatoi@post.kek.jp (Hayato Ito).

*Abstract*

We develop a lower critical field ($H_{c1}$) measurement system using the third-harmonic response of an applied AC magnetic field from a solenoid coil positioned above a superconducting sample. Parameter $H_{c1}$ is measured via detection of the third-harmonic component, which drastically changes when a vortex begins to penetrate the superconductor with temperature increase. The magnetic field locally applied to one side of the sample mimics the magnetic field within superconducting radio-frequency (SRF) cavities and prevents edge effects of the superconducting sample. With this approach, our measurement system can potentially characterize surface-engineered SRF materials such as Superconductor–Insulator–Superconductor multilayer structure (S–I–S structure). As a validation test, we measure the temperature dependence of $H_{c1}$ of two high-RRR bulk Nb samples and obtain results consistent with the literature. We also confirm that our system can apply magnetic fields of at least 120 mT at 4–5 K without any problem of heat generation of the coil. This field value is higher than those reported in previous works and makes it possible to more accurately estimate $H_{c1}$ at lower temperatures.

Keywords: Critical field $H_{c1}$, Third harmonic, SRF cavities, Thin film, S–I–S structure

# 1 Introduction

Continuous research on superconducting radio-frequency (SRF) accelerator cavities composed of Nb over the past several decades has led to increasingly improved SRF cavity performances [1]. In recent years, a maximum surface magnetic field of ~200 mT [2,3] has been reported upon the application of various surface treatments such as local grinding combined with internal surface inspection [4,5], electropolishing, subsequent annealing for hydrogen degassing, high-pressure rinsing, and low-temperature baking of the cavity [6] (see also Refs. [7,8] for newly developed surface processing technologies). However, the currently recorded maximum surface magnetic field is considered to lie close to the theoretical limit, which is called the superheating field ($H_{sh}$); here, $H_{sh}$ represents the magnetic field at which the Meissner state becomes completely unstable [9]. Therefore, it is speculated that a greater performance of SRF cavities composed of Nb cannot be realized.

Consequently, the use of the multilayer thin film structure has been proposed to address this limitation and to increase the available maximum surface magnetic field via effectively enhancing the lower critical magnetic field ($H_{c1}$) and $H_{sh}$. The multilayer thin film structure is a structure in which a superconductor layer (S) such as NbN and an insulating layer (I) are coated onto bulk Nb (S) to create an S–I–S structure [10]. In this paper, the magnetic field at which magnetic vortices start to penetrate the multilayer thin film superconductor is specifically referred to as effective $H_{c1}$. Considering the S–I–S structure, we note that a vortex avalanche can be stopped by the prevention of vortex propagation in the insulating layer. The superconductor layer needs to have a thickness of the order of the penetration depth (such that it is not considered as bulk) while also protecting the bulk Nb from the magnetic field. Upon optimization of the thickness of each layer, the SRF cavity can be made to withstand higher magnetic fields. This, in turn, means that the cavity can achieve higher accelerating field ($E_{acc}$) values than conventional SRF cavities. In this regard, the contour curve depicting the relationship between the available maximum surface magnetic field and the thickness parameters has been calculated theoretically [11], [12], [13].

To validate the theoretical prediction for the available maximum surface magnetic field, we need to measure the effective $H_{c1}$ of multilayer samples. Here, the magnetic field must be applied to one side of the sample to mimic the field distribution in SRF cavities. Otherwise, the measurement result cannot be translated to the cavity performance, for e.g., if the parallel magnetic field is applied to both sides of thin film, $H_{c1}$ is significantly enhanced, but this is never so in SRF cavities, in which the field is applied only to one side. In addition, the measurement must not be sensitive to the sample edge at which the thickness of each layer cannot be guaranteed and magnetic field is enhanced by the edge shape (edge effects). Therefore, measurements with superconducting quantum interference devices (SQUIDs), which are commonly used for magnetic property measurement, are not suitable for effective $H_{c1}$ measurement of multilayer samples except for the case that a sample has a special geometry [14]. For these reasons, the $H_{c1}$ measurement system using the third-harmonic response of an applied AC magnetic field from a solenoid coil [15] has attracted considerable attention in the context of thin film research for SRF application [16], [17]. In the third-harmonic measurement method, the magnetic field is applied on one side of a sample. If the size of the solenoid coil is considerably smaller than the sample, the sample can be considered to be an infinite plane with respect to the applied magnetic field without the influence of edge effects. Furthermore, the third-harmonic measurement method affords non-contact and non-destructive measurements of the superconductor sample.

However, in previous researches based on the third-harmonic measurement method, the applied magnetic field was limited to 40–60 mT. Since the SRF cavity operates in the temperature range of 2 to 4.2 K, a solenoid coil that can apply even higher magnetic fields is required for accurate extrapolation to the $H_{c1}$ value in the low-temperature regime [15], [16], [17]. Thus, in this study, we developed a third-harmonic $H_{c1}$ measurement system equipped with a solenoid coil that can apply significantly higher magnetic fields than those reported in previous studies, and we verified that the system functions as an $H_{c1}$ measuring device by measuring the temperature dependence of $H_{c1}$ of two high-RRR bulk Nb samples.

The paper is organized as follows. In Section 2, we present an overview our measurement system: cryostat, sample and copper stage, simulations and measurements of the magnetic field from the solenoid coil, and measurement circuit. In Sec. 3, we present the measurement results of bulk Nb samples. Subsequently, we analyze the third-harmonic signals and extract $H_{c1}(T)$ of bulk Nb samples. In Sec. 4, we discuss the validity of our measurement system.

## 2 Measurement system

### 2.1 *Measurement principle*

We consider the situation in which the solenoid coil is placed on a type-II superconductor sample and an AC magnetic field ($H_{ap}$) is applied to the sample from the coil. In this situation, a shielding current flows on the surface of the sample to prevent the penetration of magnetic flux, and the voltage in the coil is induced from both the shielding current and the AC current in the coil itself. In the state where the superconductor prevents the penetration of magnetic flux ($H_{ap} < H_{c1}$), the shielding current completely follows $H_{ap}$ and the response is linear, and thus, the third-harmonic voltage is not induced. On the other hand, in the state wherein the magnetic flux penetrates the superconductor ($H_{ap} > H_{c1}$), the response of the shielding current is saturated and nonlinear. This causes a nonlinear voltage response in the coil, thereby generating a third-harmonic voltage. In our measurement system, the sample is first cooled with a zero magnetic field. Next, the temperature of the sample is slowly raised from that of the Meissner state while applying a 1-kHz AC magnetic field from the solenoid coil to the superconducting sample. At first, since the sample is in the Meissner state, the applied magnetic field is completely expelled; however, as the temperature of the sample exceeds a specific temperature, the applied magnetic field cannot be completely expelled, and penetration of the magnetic flux occurs. Subsequently, the third-harmonic component (3 kHz) is induced in the solenoid coil. At this time, we measure the field $H_{c1}$ at this temperature by detection of the variation in the third-harmonic component. Further, we measure the temperature dependence of $H_{c1}$ by repeating the measurement with various applied magnetic fields. The selection of the 1-kHz frequency in this measurement system was determined as per a previous study [15].

### 2.2 *Cryostat*

Our experiment was carried out by cooling the sample with liquid helium (LHe) of approximately 25 L stored at the bottom of a cryostat. The cryostat consisted of an inner container and an outer tank. The inner container of the cryostat was composed of a cylinder with a diameter of 400 mm and depth of 1543 mm, and the cylinder had an extended section in the middle wherein the diameter was 610 mm and two GM refrigerators (ULVAC UR4K1040T and UR4K1060T) were mounted. LHe was supplied from a helium dewar by means of a transfer tube, and the LHe liquid level was monitored by an LHe level sensor. Before cooling down, the space in between the inner container and outer tank was pumped down for thermal insulation, and the air in the inner container was substituted with He gas. The cryostat was equipped with two GM refrigerators to reduce LHe consumption during the experiments and was precooled by these GM refrigerators before LHe transfer. After the experiment, a heater, which was installed at the bottom of the inner cylinder, was turned on for LHe to be evaporated (see Figs. 1 and 2).

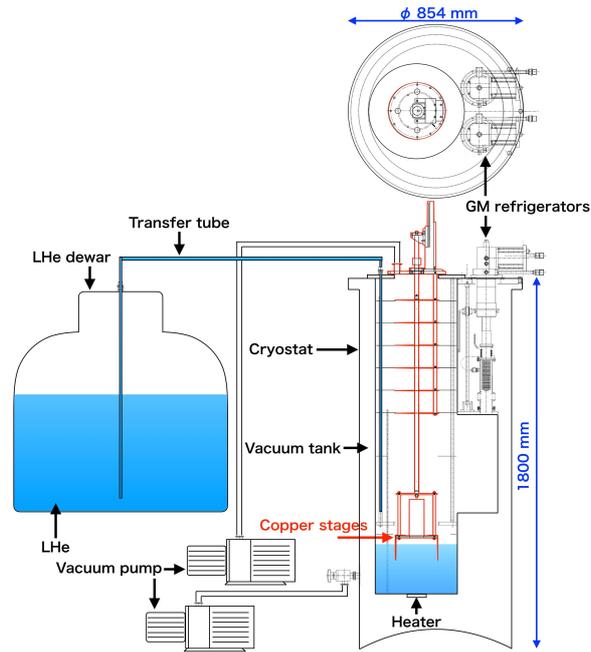

Figure 1. Schematic of cryostat used in our experiment. The red-colored components indicate the measuring equipment newly constructed for third-harmonic measurements.

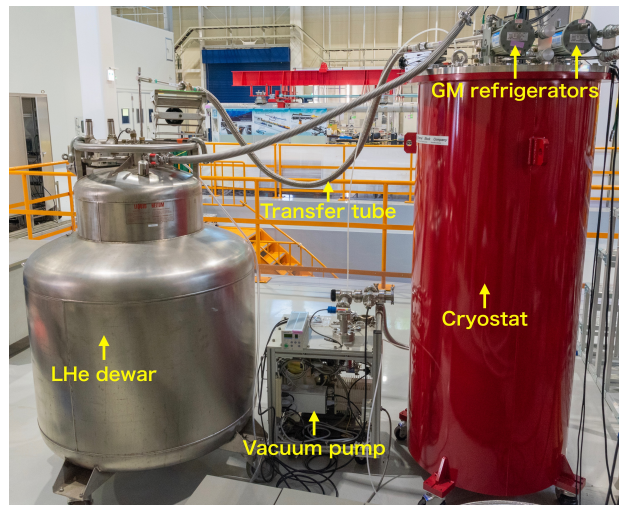

Figure 2. Photograph of complete measurement setup for experiment.

**2.3** *Copper stage*

Figure 3 shows our measurement setup, which consists of two copper disc stages, with each disc stage having a diameter of 200 mm and thickness of 5 mm. The upper stage contains the solenoid coil at the center, and the superconducting sample is positioned in between the two stages. The lower stage has two fins (width 70 mm, height 120 mm) that are extended along the upper direction, and each fin is equipped with a heater (cylinder type 110 V-80 W) to increase the temperature of the sample. The upper stage also has two fins (width 40 mm, height 80 mm) extending along the bottom direction, whose bottom ends are immersed in LHe to function as the thermal anchor and to cool down the coil. The coil stage has two slits (width 1 mm, length 37 mm) to prevent heat generation due to eddy currents (Fig. 4). In our study, we used Cernox sensors for temperature sensing. One of the temperature sensors was directly in contact with the rear surface of the sample, and the three remaining temperature sensors were used to monitor the temperature of any given part of the stage. The sample was positioned between the two stages and bolted. A gap distance of 0.05 mm between the sample surface and the solenoid coil was maintained with the use of 9 SiN balls (diameter of 3 mm) embedded in the coil stage. The solenoid coil and SiN balls were fixed to the coil stage with the use of epoxy adhesive (Araldite) (Refer also a similar setup developed in Kyoto University [18], [19]).

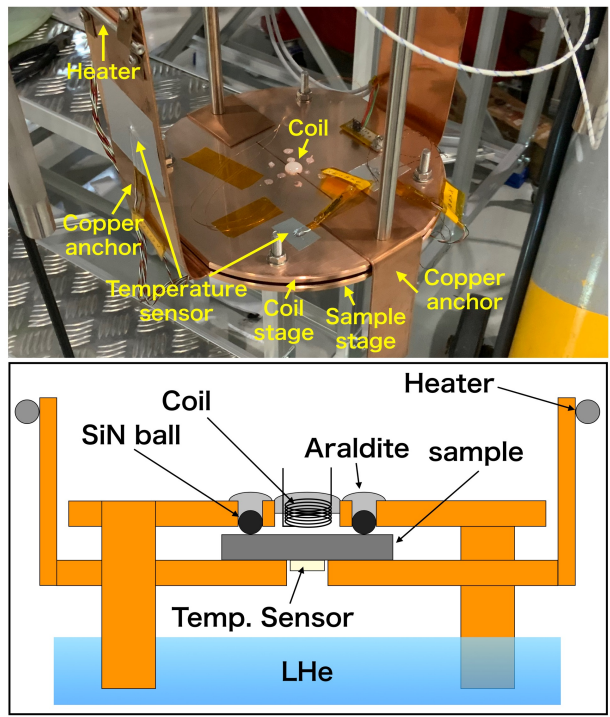

Figure 3. Copper stage setup for third-harmonic measurement. The upper photograph shows the copper stage setup. The lower panel shows the cross-sectional schematic of the copper stage setup.

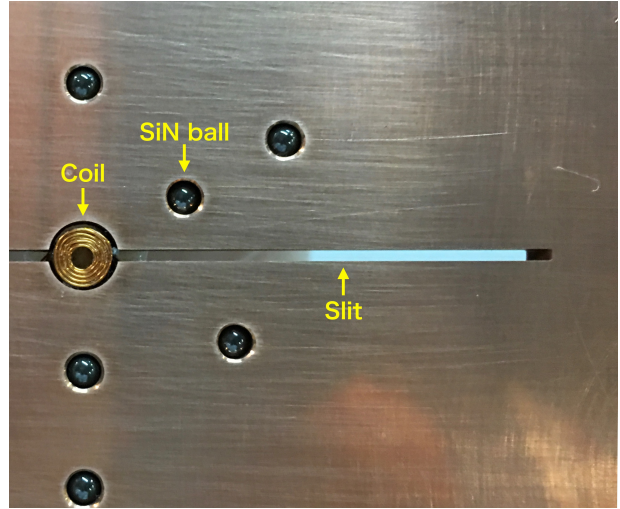

Figure 4. Photograph of coil stage on sample side.

### 2.4 *Solenoid coil*

The solenoid coil was required to be sufficiently small relative to the sample size (50 mm × 50 mm). To determine the size of the solenoid coil, we performed magnetic field simulations using Finite Element Method Magnetics (FEMM) [20]. The solenoid coil was positioned 0.05 mm above the sample boundary such that the actual setup was reflected in the geometry of the model. Figure 5 shows the simulated magnetic field profile of the solenoid coil based on the axisymmetric model and magnetic field strength of the sample surface along the radial direction. As regards the simulation result, the peak value of the magnetic field was determined as 111 mT with a current of 4.5 A upon assuming that the inner diameter of the coil was 2 mm, outer diameter was 5 mm, length was 5 mm, and number of turns was 176 (these values are the actual coil parameters reflecting the actual setup). In addition, the magnetic field at the sample edge (25 mm) was 0.4% of the peak, and thus, we considered the sample edge effect to be negligible. To confirm the validity of the magnetic field simulations based on FEMM, we measured the magnetic field at a point closest to the solenoid coil using a gauss meter (F. W. BELL Hand-held Gauss/Tesla Meter Model 4048) and compared it with the calculated result. Since the sensor thickness of the gauss meter was 1.3 mm, the magnetic field distribution at a distance of 0.65 mm from the solenoid coil was simulated and averaged using the cross-sectional area (2 mm × 3 mm) of the sensor. Figure 6 compares the measured magnetic field and the calculated result; there is a difference of 2.4% between the measured magnetic field and the calculated result. Therefore, the difference of 2.4% is taken into account as the correction factor for calculation of the applied magnetic field on the surface in our measurement. Consequently, the relationship between the magnetic field on the surface and the (corrected) coil current from the simulation result provides information on the applied magnetic field in our measurement system (Fig. 7).

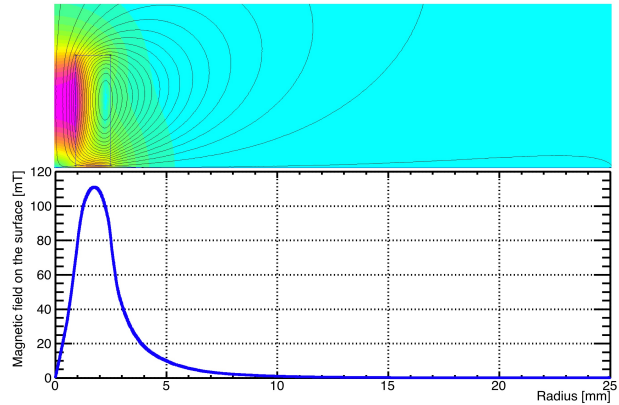

Figure 5. Magnetic field profile of solenoid coil as simulated by Finite Element Method Magnetics (FEMM). The upper panel depicts the magnetic field lines along the radial cross-section. The lower panel depicts the magnetic field strength on the sample surface along the radial direction.

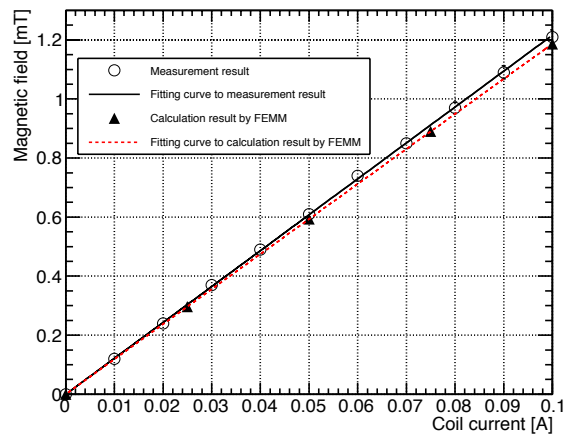

Figure 6. Comparison of measured magnetic field and calculated result. The difference of 2.4% between the black solid curve and the red dashed curve is taken as a correction factor in the magnetic field calibration.

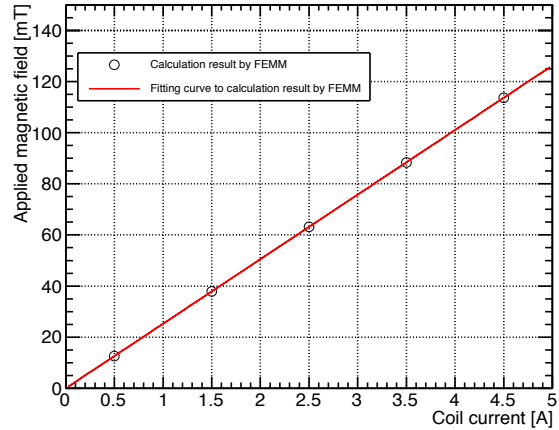

Figure 7. Relationship between magnetic field on surface and coil current after correction. The open circles indicate the calculated magnetic field values on surface. The peak value of the magnetic field was corrected to 114 mT for a current of 4.5 A. The red curve represents the fitting result to the linear function.

**2.5** *Measurement circuit*

Figure 8 shows the block diagram of the measurement circuit. In this setup, the signal generator (TEXIO FGX-2005) generates a 1-kHz sinusoidal waveform with an amplitude of 1 $V_{pp}$. The 1-kHz signal is filtered by a 1-kHz bandpass filter (BPF) with a bandwidth of ±20 Hz and applied to an amplifier (max. output of 20 $V_{pp}$ for 2-Ω termination). The amplified 1-kHz signal is applied to the solenoid coil in the cryostat through the coaxial cable and the BNC-type feedthrough to generate the magnetic field applied to the sample. The amplified 1-kHz signal is detected at both ends of the coil and fed into a 3-kHz BPF with a bandwidth of ±30 Hz through the coaxial cable and the BNC-type feedthrough. The detected 3-kHz signal is amplified with selectable gain values of 10, 100, and 1000. The waveform of the detected 3-kHz signal is acquired by means of an oscilloscope (PicoScope 4262, 16-bit resolution), and the fast Fourier transform (FFT) is performed to measure the signal strength. The current is monitored via measuring the voltage across the 0.1-Ω resistor. In the study, we set a 50% duty pulse operation (on-time 2 s, off-time 2 s) to reduce coil heating. In this measurement system, the temperature data of the four temperature sensors, signal strength and phase of the detected 3-kHz signal, and voltage across the 0.1-Ω resistor were recorded by the computer as one data set during the on-time of 2 s.

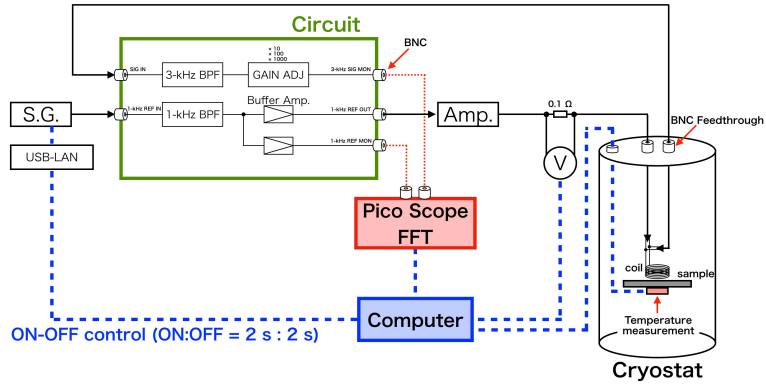

Figure 8. Block diagram of measurement circuit.

## 3 Measurement results

In order to verify that our proposed device functions as an $H_{c1}$ measuring probe for type-II superconductors and that a high magnetic field of ~150 mT can be applied to the sample, we tested two bulk Nb samples from the same Nb-plate production lot. The Nb plates used in this research were manufactured by Tokyo Denkai Co., Ltd. with a residual resistivity ratio (RRR) >300 and thickness of 2.8 mm. After cutting and drilling a hole in the sample, except for low-temperature baking, the same process as that applied to the SRF cavity was applied to the Nb sample (surface grinding with sponge abrasive, bulk electropolishing of 100 μm, annealing for 4 h at 750 °C for hydrogen degassing, light electropolishing of 20 μm, ultrasonic cleaning with detergent FM 20). The size of the flat surface of the sample was 50 mm × 50 mm.

### 3.1 *Signal*

The cooling down of the Nb sample to the Meissner state was performed without any application of the magnetic field, i.e., with zero magnetic field. The measurement was performed by fixing the applied AC current to the solenoid coil and slowly increasing the temperature. In this measurement, the heater output power was manually adjusted to maintain the temperature increase rate to 0.1 K/min or less to obtain a uniform increase in the temperature of the whole sample. Figure 9 shows the profile of the temperature in the measurement. The rate of temperature rise was calculated as 0.048 K/min from the slope of the red curve in Fig. 9. Figure 10 shows the typical third-harmonic response vs. the temperature for the bulk Nb sample. A significant change in the third-harmonic signal is detected around 5.8 K when the applied AC current is 3.8 A. Ideally, the third-harmonic signal should remain zero until the first vortex penetrates the sample. In our setup, the AC magnetic field generated by the solenoid coil distorts the solenoid coil itself, and thus, the third-harmonic signal exhibits a DC offset. Next, we analyzed the third-harmonic signal strength to determine the temperature at which the vortex starts to penetrate the sample. Linear fitting was applied to the DC offset in the temperature range lower than the left onset, at which the third-harmonic signal starts to change drastically. Subsequently, the first point beyond the $3\sigma$ value determined from the distribution of the difference between each point and the linear fitting function was taken as the temperature at which the vortex penetrates the sample (penetration temperature). Because the variation in the penetration temperature for each measurement for the same value of the applied magnetic field is at most 0.1 K, the temperature error in each penetration temperature was uniformly determined to be 0.1 K. The error of the applied magnetic field was determined from the deviation of the current value at each measurement point within this temperature error.

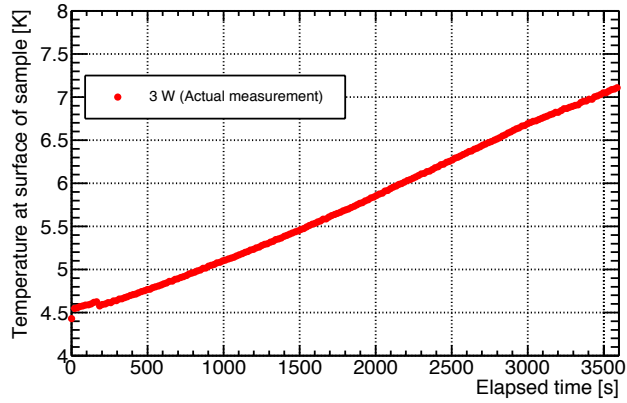

Figure 9. Plot of temperature of sample vs. elapsed time during sample measurement at around 3 W with heater power control. The heater power was adjusted to 3 W at 200 s.

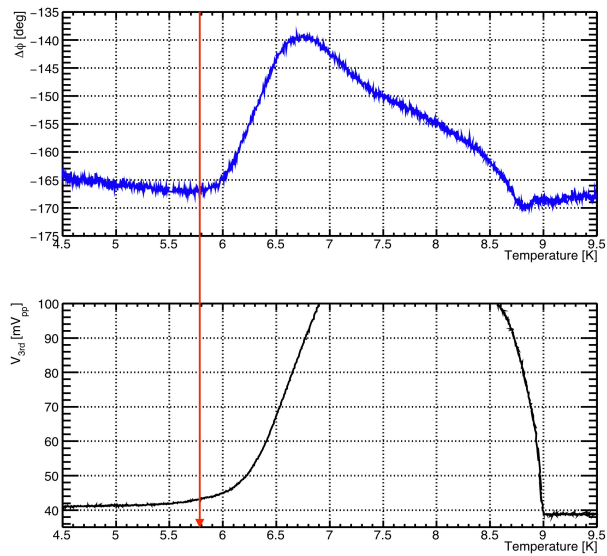

Figure 10. Typical third-harmonic signal of bulk Nb sample 1. The upper graph shows the phase variation of the third harmonic signal, while the lower graph shows the variation in the third-harmonic signal strength. Note that the third-harmonic signal strength is amplified by a factor of 10 by means of the selectable gain adjustor. The red arrow represents the onset of the third-harmonic signal as per our analysis of the third-harmonic signal strength.

### 3.2 *Temperature dependence of $H_{c1}$*

The above measurement and analysis were performed repeatedly at various values of the applied magnetic field. The magnetic field was calibrated by converting the current applied to the solenoid coil into the corresponding magnetic field using the FEMM simulation results. In Fig. 11, the red and blue lines depict the resultant temperature dependence of $H_{c1}$ of the two samples, which were obtained by fitting the empirical formula $H_{c1}(T) = H_{c1}(0)[1 - (T/T_c)^2]$ with the measured data. The $T_c$ values were calculated to be 9.08 ± 0.06 K and 8.98 ± 0.05 K for Nb samples 1 and 2, respectively, via calculating the fitting value at the temperature intercept. Further, the $H_{c1}(0)$ values were calculated as 167 ± 2 mT and 174 ± 2 mT for Nb samples 1 and 2, respectively, by extrapolating the fitting curve to the temperature of 0 K.

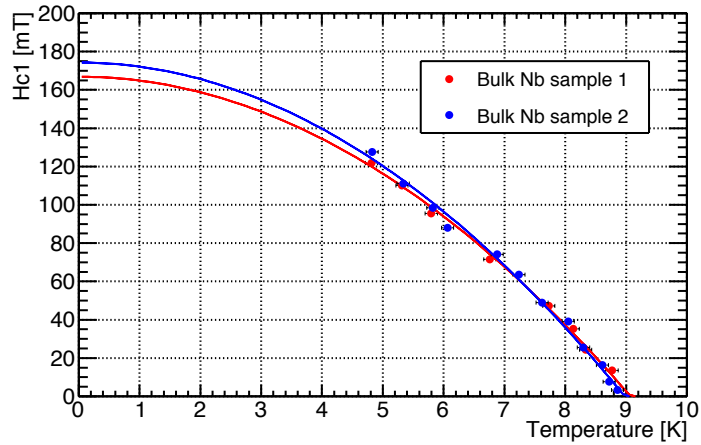

Figure 11. Temperature dependence of $H_{c1}$ of two bulk Nb samples.

## 4 Discussion

In this measurement, a magnetic field at around 130 mT was applied to the sample at currents of up to 5 A without any problem of heat generation of the coil. Here, we note that the LHe temperature of 4.2 K determines the lower limit of the measurable temperature in Fig. 11, and an even higher magnetic field can be applied in our measurement system. If heat generation of the coil is not taken into account, a value double the magnetic field can be applied in principle since the amplifier limit is 10 A. The reliability of our system is supported by the measurement result because the average value of 171 ± 2 mT at 0 K is reasonable when compared with the reported $H_{c1}$ values (173.5 mT, 185 mT) from the references [21], [22].

It is also interesting to compare our measurement result with $E_{acc}$ of SRF cavities, which is thought to be limited by strong vortex dissipation at $H_{c1} < H < H_{sh}$. Converting the average $H_{c1}$ value at 2 K (162 ± 2 mT) in this study into the corresponding $E_{acc}$ value in the SRF cavity (TESLA cavity) considering the operation temperature of the SRF cavity results in an $E_{acc}$ value of 38.1 ± 0.4 MV/m. This value is consistent with the distribution of the maximum $E_{acc}$ performance obtained from RF test of the 9-cell TESLA cavity [23].

Since our measurement system can potentially apply high magnetic fields >120 mT to the sample and since the $H_{c1}$ value of the bulk Nb can be measured to an accuracy of 1% in our experiments, the $H_{c1}$ measurement of thin-film samples with high $H_{c1}$ values, such as those of the S–I–S structure, can be performed with a relative value based on the $H_{c1}$ value of bulk Nb, and this difference can be measured with an accuracy of 1%.

## 5 Summary

In this paper, we described in detail the construction of a third-harmonic-based $H_{c1}$ measurement system and our $H_{c1}$ measurement results for two high-RRR bulk Nb samples. The geometry of the solenoid coil was optimized via FEMM to apply a high magnetic field of above 150 mT to the sample and overcome sample edge effects. The $H_{c1}$ values of two bulk Nb samples, whose surfaces were treated with the same process as that applied to the SRF cavity except for low-temperature baking, were measured as a verification test. The measurement was performed via slowly increasing (below the rate of 0.1 K/min) the sample temperature. A significant change in the third-harmonic response was detected at up to the LHe temperature, which is the limit of the measurable temperature in our system. As regards the measurement results, we obtained an average $H_{c1}(0)$ value of bulk Nb as 171 ± 2 mT, which is consistent with the literature, and we found that a magnetic field of at least 120 mT can be applied to the sample without any problem of heat generation of the coil. This result suggests that thin-film samples can be measured and compared accurately with bulk Nb measurements by use of our measurement system.

**Acknowledgement**

The authors would like to express their sincere gratitude to Dr. C. Z. Antoine (CEA, Irfu), Dr. Y. Iwashita (Kyoto University), and Dr. R. Katayama (KEK) for their useful discussions on the development of our measurement system.

**Funding**: This work was supported by the Japan Society for the Promotion of Science Grant-in-Aid for Young Scientist (A) No. 17H04839.